\documentclass[12pt]{emulateapj}
\renewcommand{\deg}{\mbox{$^{\circ}$}}
\def\farcm{\hbox{$.\mkern-4mu^\prime$}}
\def\farcs{\hbox{$^{\prime\prime}$}}

\def\ltsim{\, {}^<_\sim \,}

\def\deg{\ifmmode^\circ\else$^\circ$\fi}    %Degree sign%

\def\hper{\ifmmode \rlap.^{h}\else $\rlap{.}^h$\fi} 
\def\sper{\ifmmode \rlap.^{s}\else $\rlap{.}^s$\fi}    %Superscript 's' over per
\def\Deg{${}^\circ$\llap{.}}

\def\Sec{${}^{\prime\prime}$\llap{.}}
\def\deg{${}^\circ$}

\def\bmv{\hbox{\em B--V\/}}
\def\bmi{\hbox{\em B--I\/}}

\def\today{\number\year\space \ifcase\month\or
  January\or February\or March\or April\or May\or June\or
  July\or August\or September\or October\or November\or December\fi
  \space\number\day}
\def\now{\number\year\space \ifcase\month\or
  January\or February\or March\or April\or May\or June\or
  July\or August\or September\or October\or November\or December\fi
  \space\number\day .\number\time}

\shortauthors{Di Cecco et al.}

\begin{document}
\title{On the absolute age of the Globular Cluster M92\altaffilmark{1}}

\author{
A.\ Di Cecco\altaffilmark{2}, 
R.\ Becucci\altaffilmark{3},
G.\ Bono\altaffilmark{2,4},
M.\ Monelli\altaffilmark{5},
P.\ B.\ Stetson\altaffilmark{6,13,14},
S.\ Degl'Innocenti\altaffilmark{3,7}, 
P.\ G.\ Prada Moroni\altaffilmark{3,7},
M.\ Nonino\altaffilmark{8},
A.\ Weiss\altaffilmark{9},
R.\ Buonanno\altaffilmark{2,10},
A.\ Calamida\altaffilmark{11},
F.\ Caputo\altaffilmark{4},
C.\ E.\ Corsi\altaffilmark{4},
I.\ Ferraro\altaffilmark{4},
G.\ Iannicola\altaffilmark{4},
L.\ Pulone\altaffilmark{4},
M.\ Romaniello\altaffilmark{11}, and
A.\ R.\ Walker\altaffilmark{12}}

\altaffiltext{1}{This paper makes use of data obtained from the Isaac Newton Group
Archive which is maintained as part of the CASU Astronomical Data Centre 
at the Institute of Astronomy, Cambridge. This research used the facilities of the Canadian 
Astronomy Data Centre operated by the National Research Council of Canada with the support of the Canadian Space Agency. 
}
\altaffiltext{2}{Dipartimento di Fisica, Universit\`a di Roma Tor Vergata, via della 
Ricerca Scientifica 1, 00133 Rome, Italy; alessandra.dicecco@roma2.infn.it}
\altaffiltext{3}{Dipartimento di Fisica, Universit\`a di Pisa, Largo B. Pontecorvo 2, 56127 Pisa, Italy}
\altaffiltext{4}{INAF--OAR, via Frascati 33, Monte Porzio Catone, Rome, Italy}
\altaffiltext{5}{IAC, Calle Via Lactea, E38200 La Laguna, Tenerife, Spain}
\altaffiltext{6}{DAO--HIA, NRC, 5071 West Saanich Road, Victoria, BC V9E 2E7, Canada}
\altaffiltext{7}{INFN--Pisa, via E. Fermi 2, 56127 Pisa, Italy}
\altaffiltext{8}{INAF--OAT, via G.B. Tiepolo 11, 40131 Trieste, Italy}
\altaffiltext{9}{Max-Planck-Institut für Astrophysik, Karl-Schwarzschild-Str. 1, 85748 Garching, German}
\altaffiltext{10}{ASI--Science Data Center, ASDC c/o ESRIN, via G. Galilei, 00044 Frascati, Italy}
\altaffiltext{11}{ESO, Karl-Schwarzschild-Str. 2, 85748 Garching bei Munchen, Germany}
\altaffiltext{12}{CTIO--NOAO, Casilla 603, La Serena, Chile}
\altaffiltext{13}{Visiting Astronomer, Kitt Peak National Observatory, National Optical Astronomy 
Observatory, which is operated by the Association of Universities for Research in Astronomy (AURA) 
under cooperative agreement with the National Science Foundation.}
\altaffiltext{14}{Visiting Astronomer, Canada-France-Hawaii Telescope operated by the
National Research Council of Canada, the Centre National de la Recherche
Scientifique de France and the University of Hawaii.}

\date{\centering drafted \today\ / Received / Accepted }

\begin{abstract}
We present precise and deep optical photometry of the globular M92. 
Data were collected in three different photometric systems: Sloan Digital Sky 
Survey ($g'$,$r'$,$i'$,$z'$; MegaCam@CFHT), Johnson-Kron-Cousins ($B$,$V$,$I$; 
various ground-based telescopes) and Advanced Camera for Surveys (ACS) Vegamag ($F475W$, 
$F555W$, $F814W$; Hubble Space Telescope). Special attention was given to 
the photometric calibration, and the precision of the ground-based data is generally 
better than 0.01 mag. We computed a new set of $\alpha$-enhanced evolutionary models 
accounting for the gravitational settling of heavy elements at fixed chemical composition 
([$\alpha$/Fe]=+0.3, [Fe/H]=--2.32 dex, $Y$=0.248). The isochrones---assuming the 
same true distance modulus ($\mu$=14.74 mag), the same reddening ($E(\bmv)$=0.025$\pm$0.010 mag),
and the same reddening law---account for the stellar distribution along the 
main sequence and the red giant branch in different Color-Magnitude Diagrams 
($i'$,$g'-i'$; $i'$,$g'-r'$; $i'$,$g'-z'$; $I$,$B-I$; $F814W$,$F475W-F814W$). 
The same outcome applies to the comparison between the predicted 
Zero-Age-Horizontal-Branch (ZAHB) and the HB stars. We also found 
a cluster age of 11$\pm$1.5~Gyr, in good agreement with previous estimates. 
The error budget accounts for uncertainties in the input physics and the photometry.      
To test the possible occurrence of CNO-enhanced stars, we also computed two 
sets of $\alpha$- and CNO-enhanced (by a factor of three) models both 
at fixed total metallicity ([M/H]=--2.10 dex) and at fixed iron abundance. We found that 
the isochrones based on the former set give the same cluster age (11$\pm$1.5~Gyr) as the 
canonical $\alpha$-enhanced isochrones.  The isochrones based on the latter set also 
give a similar cluster age (10$\pm$1.5~Gyr). These findings support previous results 
concerning the weak sensitivity of cluster isochrones to CNO-enhanced chemical 
mixtures.   
\end{abstract}


\maketitle

%%%%%%%%%%%%%%%%%%%%%%%%%%%%%%%%%%%%%%%%%%%%%%%%%%%%%%%%%%%%%%%%%%%%%
\section{Introduction}

The Galactic globular cluster (GGC) M92 (NGC~6341) is among the oldest and 
most metal-poor ([Fe/H]=--2.38, Kraft \& Ivans 2003; 2004) Galactic stellar 
systems.  It is located well above the Galactic plane (l=68.34$^{\circ}$, 
b=34.86$^{\circ}$, Harris 1996), is minimally affected by field 
star contamination, and is only slightly reddened (E(\bmv)=0.02, Harris 1996). 
However, the cluster's absolute age is far from being well established. Recent estimates 
based on updated cluster isochrones indicate that the absolute age of M92 ranges from 
$12.3\pm0.9$~Gyr (Salaris \& Weiss \ 2002) to 14.8$\pm$2.5 Gyr (Carretta et al.\ 2000; 
Grundahl et al.\ 2000). 
A very accurate estimate of the absolute age of M92 was also provided by VandenBerg et al.\ (2002). 
They gave a new estimate of both distance modulus (DM) and reddening ($DM_V$=14.60$\pm$0.12, 
E(\bmv)=0.023 mag) 
using the unique field, metal-poor ([Fe/H]$\lesssim$--2.3) subdwarf calibrator with an accurate 
trigonometric parallax from {\em Hipparcos} ($\sigma_\pi/\pi \le$0.1). By adopting the quoted
values and a set of cluster isochrones ([Fe/H]=--2.3, [$\alpha$/Fe]=0.3 dex) including gravitational 
settling and radiative accelerations they found an absolute age of 13.5$\pm$1.0-1.5 Gyr.   
Similar cluster ages have been found using the luminosity 
function (LF) of evolved stars, and indeed Cho \& Lee (2007) found an age of 
$\sim$13.5~Gyr, while Paust, Chaboyer \& Sarajedini \ (2007) found an age of 14.2$\pm$1.2~Gyr.
Precise estimates of the GC absolute ages rely, from an empirical point of view, 
on four ingredients: distance, total metallicity, reddening, and photometric 
zero-points (Renzini\ 1991). 
Recent estimates of the cluster distance modulus based on different standard 
candles agree quite well. The true DM ($\mu$) of M92 based on main sequence (MS) 
fitting  ranges from $\mu=14.64\pm 0.07$ mag (Carretta et al.\ 2000) 
to $\mu=14.75\pm 0.11$ mag (Kraft \& Ivans 2003).
Distances based on the near-infrared (NIR) Period-Luminosity (PL) relation 
of RR Lyrae stars (Bono et al.\ 2001; Cassisi et al.\ 2004; Catelan 2004; 
Del Principe et al.\ 2005, 2006; Sollima et al.\ 2006) suggest a similar mean 
value, i.e., $\mu$=14.65$\pm$0.10 mag. Thus, the different distance estimates 
agree within their current uncertainties (1$\sigma$). 

The uncertainty in the reddening correction for low-reddening GCs is typically 
of the order of 0.02 mag (Gratton et al.\ 2003; Bono et al.\ 2008 ).  Experience
has shown that the typical uncertainty in the zero-point of photometry from a
given observing run is still of the order of 0.01--0.02 mag (e.g., Stetson, Bruntt \&
Grundahl 2003; Stetson, McClure \& VandenBerg 2004; Stetson 2005).  In the case of M92,
this uncertainty may occasionally have been still larger (0.03 mag, Carretta et al.\ 2000).
However, we believe that this systematic uncertainty can be reduced by averaging the
photometric results of many observing runs, each of which has been individually
calibrated to a common photometric standard system.  The uncertainty affecting 
the total metallicity is of the order of 0.2 dex, if we include
cluster-to-cluster uncertainties in iron and $\alpha$-element abundances as well as systematic
uncertainties in the overall metallicity scale (Carretta et al.\ 2009).    

Dating back to the seventies, spectroscopic measurements showed quite clearly the occurrence of 
star-to-star variations of C and N abundances in several GGCs (M5, M10, M13, M92, $\omega$ Cen, Osborn 1971;
Cohen 1978). Subsequent investigations also found variations in Na (Cohen 1978; Peterson 1980), 
in Al (Norris et al.\ 1981) and in O (Pilachowski et al.\ 1983; Leep, Wallerstein \& Oke \ 1986).
The observational scenario was further enriched by the evidence that the molecular band-stregths 
of CN and CH appear to be anticorrelated (Smith 1987; Kraft 1994, and references therein).  
Both this anticorrelation and an anticorrelation between O--Na and Mg--Al have 
been observed in evolved (red giant branch [RGB], Horizontal Branch [HB]), and in 
unevolved MS stars of all GCs studied in sufficient detail (Suntzeff \& Smith 1991; Cannon et al.\ 1998; 
Harbeck, Grebel \& Smith 2003; Gratton et al.\ 2001; Ramirez \& Cohen 2002; Carretta et al.\ 2007).

A working hypothesis to explain these observations is that a previous generation (first 
generation) of asymptotic giant branch (AGB) stars expelled processed material during thermal 
pulses, thus the subsequent stellar generation (second generation) formed with 
material that had already been polluted (Ventura et al.\ 2001; Gratton, Sneden \& Carretta \ 2004, and references therein). 
In this scenario, the surface abundance of the second stellar generation is characterized by 
a significant He enrichment and by well defined C-N-O-Na anticorrelations. It has been suggested 
that typically the fraction of stars belonging to the second generation might be of order 
50\% (D'Antona \& Caloi 2008).    
Alternative hypotheses are that cluster self-pollution is caused either by evolved RGB stars 
that experienced extra-deep mixing (Denissenkov \& Weiss 2004) or by fast rotating intermediate-mass 
stars (Maeder \& Meynet 2006; Prantzos \& Charbonnel 2006; Decressin et al.\ 2007). 
M92 shows the typical variations in [C/Fe] and [N/Fe] 
(Carbon et al.\ 1982; Langer et al. 1986;  Bellman et al. 2001), together with the usual anticorrelations 
(Pilachowski 1988; Sneden et al.\ 1991; Kraft 1994). However, up to now no evidence has 
been found concerning the occurrence of multiple stellar populations (Piotto et al.\ 2007; 
Cassisi et al.\ 2008). Therefore, M92 is a perfect laboratory to constrain the impact of 
canonical and CNO-enhanced mixtures on the estimate of the cluster age.    

We take advantage of deep and accurate multiband optical images collected with both 
ground-based and space telescopes to constrain the absolute age of M92.

%%%%%%%%%%%%%%%%%%%%%%%%%%%%%%%%%%%%%%%%%%%%%%%%%%%%%%%%%%%%%%%%%%%%%%%%%%%%%%%%%%%%%
% 			fig 1
%%%%%%%%%%%%%%%%%%%%%%%%%%%%%%%%%%%%%%%%%%%%%%%%%%%%%%%%%%%%%%%%%%%%%%%%%%%%%%%%%%%%%
\begin{figure*}[t]
\begin{center}
\label{fig1}
\includegraphics[height=8cm,width=14cm]{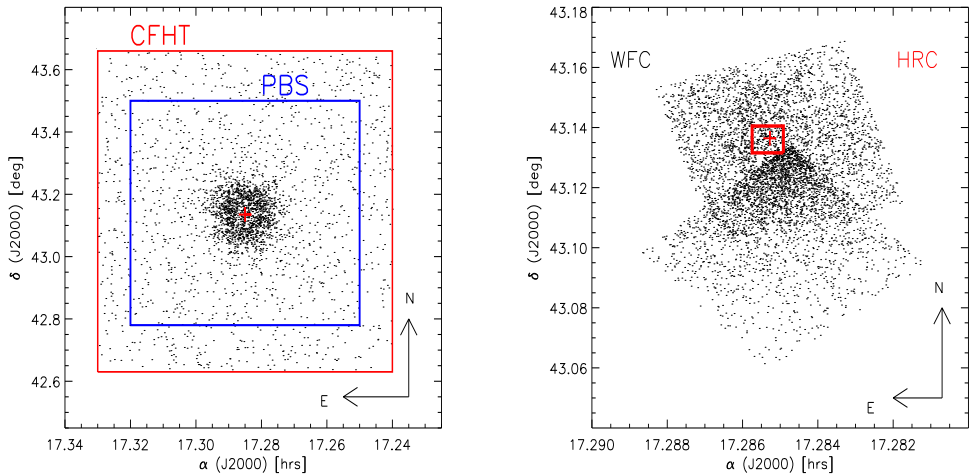}
\vspace*{-0.15truecm}
\caption{Left -- sky area across the globular cluster M92 covered by the different ground-based 
data sets. The red box shows the area covered by CFHT images, while the blue box shows 
the area covered by the Johnson-Cousins images. The orientation is shown in the bottom  
right corner. 
Right -- same as the left, but for space data sets collected with the Advanced Camera 
for Surveys (ACS) on board the HST. The dots display the area covered by the images 
collected with the Wide Field Channel (WFC),  while the red square those collected 
with the High Resolution Channel (HRC). The red cross marks the cluster center.
}
\end{center}
\end{figure*}
 
%-------------------------------------------------------------------------
%%%%%%%%%%%%%%%%%%%%%%%%%%%%%%%%%%%%%%%%%%%%%%%%%%%%%%%%%%%%%%%%%%%%%%%%%%%%%%%%%%%%%
% 			fig 2
%%%%%%%%%%%%%%%%%%%%%%%%%%%%%%%%%%%%%%%%%%%%%%%%%%%%%%%%%%%%%%%%%%%%%%%%%%%%%%%%%%%%%
\begin{figure*}[t]
\begin{center}
\label{fig2}
\includegraphics[height=13cm,width=15cm]{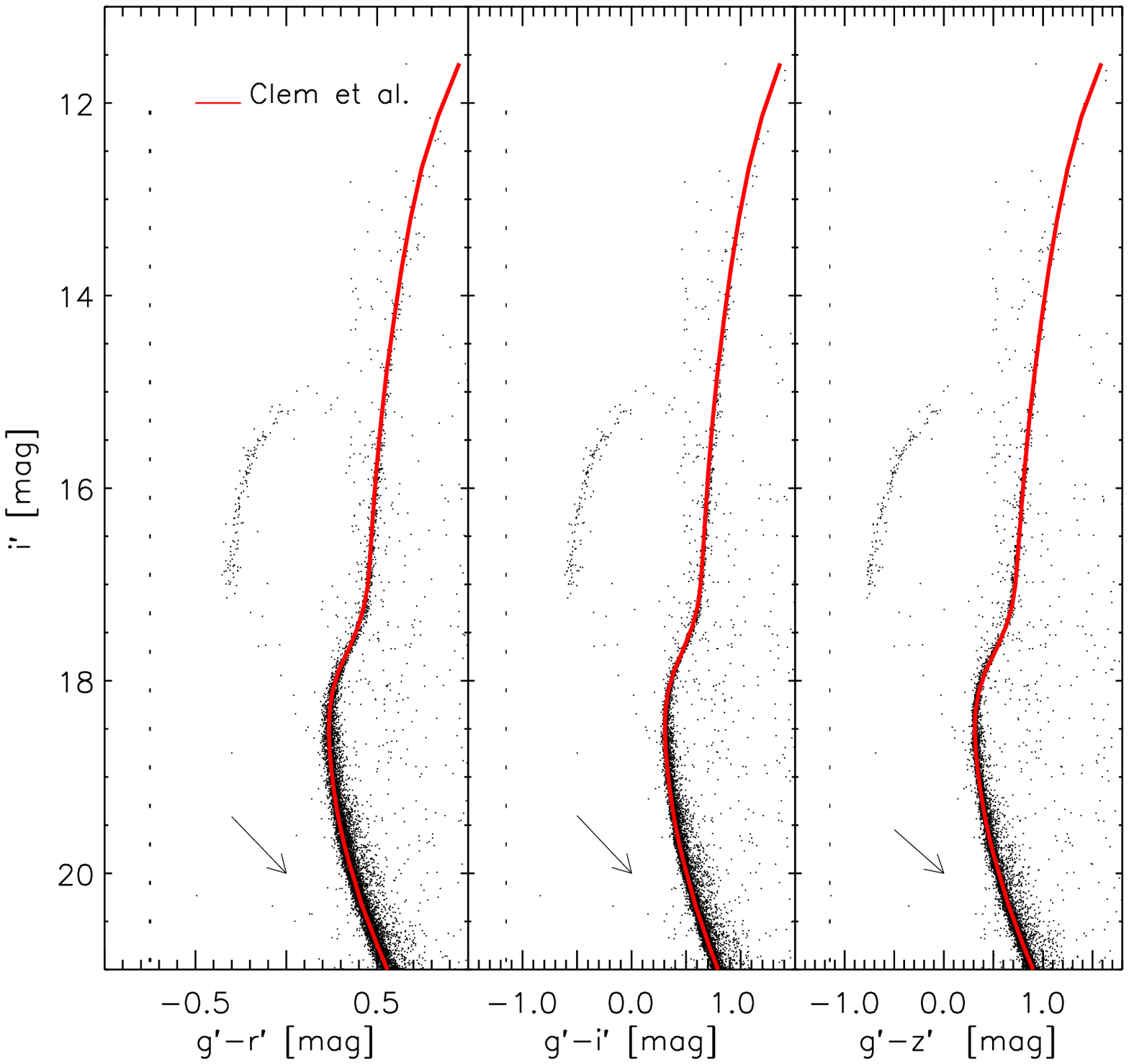}
%\vspace*{-3.25truecm}
\caption{The $i'$,$g'-r'$ (left), $i'$,$g'-i'$ (middle) and $i'$,$g'-z'$  (right) CMD based 
on images collected with the MegaCam@CFHT. The red lines display the ridgelines provided by 
Clem et al.\ (2007). The error bars on the left show the mean intrinsic photometric error 
in magnitude and in color, while the black arrows the reddening vectors.   
}
\end{center}
\end{figure*}

%%%%%%%%%%%%%%%%%%%%%%%%%%%%%%%%%%%%%%%%%%%%%%%%%%%%%%%%%%%%%%%%%%%%%%%%%%%%%%%%
\section{Data reduction and theoretical framework}

To provide robust estimates of the absolute age of M92, we secured optical images 
collected in three different photometric systems. In particular, we adopted both 
ground-based and space images. The ground-based data were collected either with 
MegaCam --- the mosaic camera available at the Canada-France-Hawaii Telescope (CFHT: Sloan
filters, field of view [FoV]: $1^\circ$x$1^\circ$,
spatial resolution: 0\farcs.19/pixel) --- or with several small/medium telescopes 
(Johnson-Cousins bands). The space data were collected with the Advanced 
Camera for Surveys (ACS) on the Hubble Space Telescope (HST) using both 
the Wide Field Channel (WFC,  FoV: 202\farcs x202\farcs, 0\farcs.05/pixel) and the High 
Resolution Channel (HRC, FoV: $0'.5$x$0'.5$,  0\farcs.025/pixel). 
The sky coverage of the different data sets is plotted in Fig. 1.
 
The CFHT data set includes 1440 CCD images collected in the Sloan bands 
--$g',r',i',z'$--  with different exposure times (ETs). In particular, the 
shallow images were acquired with ETs of 5  ($g',r',i'$) and 15 ($z'$) s, while  
the deep images had ETs of 250 ($g',r'$), 300 ($i'$) and 500 ($z'$) s. 
The mean seeing in the different bands ranges from $\sim$ 0\Sec75 ($g',r',i'$) 
to $\sim$ 0\Sec85 ($z'$). These images were pre-processed using Elixir 
(Magnier \& Cuillandre 2004). 
For each chip of the mosaic camera we performed standard Point Spread Function (PSF) fitting photometry with
DAOPHOT~IV and ALLSTAR (Stetson 1987). The individual chips were rescaled to a 
common geometrical systems defined by a 1\Deg5$\times$1\Deg5 Sloan Digital Sky Service (SDSS) 
reference image using DAOMATCH/DAOMASTER. We performed the final photometry by running ALLFRAME 
(Stetson 1994) simultaneously over the entire data set. We ended up with a photometric catalog 
including $\sim$84,000 stars with at least one measurement in two different photometric bands. 
The absolute calibration was performed using local secondary standards provided by 
Clem, VandenBerg \& Stetson \ (2007). Fig. 2 shows three different Color-Magnitude Diagrams (CMDs) 
based on this catalog. Data plotted 
in this figure were selected according to photometric error $\le 0.01$ mag, 
sharpness\footnote{The sharpness is an index that quantify the similarity between the shape 
of the measured objects and of the adopted PSF.} \textit{abs(sha)}$\le$1, 
and separation\footnote{The separation index quantifies the degree of crowding (Stetson et al.\ 2003)}
 \textit{sep}$\ge$2.5. Moreover, to overcome the central crowding and contamination
by field stars we selected stars in an annulus 60\farcs$\le r \le 700$\farcs. 
Special attention was paied to the absolute and relative photometric zero-points. 
The precision for a single star is typically better than 0.02 mag. The precision
of the different calibrations is supported by the very good agreement between
the ridge lines (red lines in Fig. 2) provided by Clem et al. and our
measurements of the cluster stars.  The error budget accounting for intrinsic
and calibration errors is smaller than 0.01 mag down to Main Sequence Turn-Off 
(MSTO) stars ($i'$$\sim$19 mag).     

In addition to the Sloan data, we also analyzed 782 CCD images collected in the
Johnson-Cousins bands; these were reduced and calibrated by one of us (PBS) in
an ongoing effort to provide homogeneous photometry on the Landolt (1992)
photometric system\footnote {For more details see the following URL:
http://www4.cadc-ccda.hia-iha.nrc-cnrc.gc.ca/ community/ STETSON/ standards/}. 
These data were obtained in the course of 44 independent observing runs on nine
telescopes (CFHT; DAO; INT; JKT; KPNO 0.9, 2.1, 4.0; NOT) from 1984 to 2002.
Based upon frame-to-frame repeatability, we infer that at least some stars have magnitudes
and colors individually precise to $\ltsim 0.01\,$mag as faint as $V\sim 20$,
which is $\sim 1.5\,$mag fainter than the MSTO.  Given the large
number of independently calibrated observing runs, we believe that systematic calibration
uncertainties should be well under 0.01$\,$mag.
  
The ground-based images were supplemented with space images collected with ACS@HST.
In particular, we used twelve images in two different pointings acquired with the 
WFC (see right panel of Fig. 1). The innermost WFC images\footnote{ GO-9453, PI: T. Brown} 
were three with the $F814W$-band (ETs of 0.5, 6, and 100 s) and covering the center of the cluster. 
%and three with the $F606W$ bands (ETs of 0.5, 5, and 90 s), covering the center of the cluster. 
The outermost WFC images\footnote{GO-10505, PI: C. Gallart} were:  three with the $F475W$ 
(ETs of 3, 20, 40 s) and three with the $F814W$ (ETs of 1, 10, 20 s) filters, located 
$\sim 2'$ S--E from the cluster center.  We used the FLT versions of these images and did 
not apply any cosmic ray mask, since the individual ETs are short.   
To overcome the crowding of the innermost regions (concentration \textit{c}=1.81, Harris 1996), 
we also considered a set of images collected with the HRC (see right panel of Fig. 1). This
data set includes 155 $F555W$ drizzled images collected with different exposure times (74$\times$10 s, 
4$\times$120 s, 10$\times$40 s, 4$\times$80 s, 34$\times$100 s, 8$\times$200 s, 4$\times$400 s,  
9$\times$500 s, 8$\times$1000 s)\footnote{GO-10335, PI: H. Ford}. 
%and six $F435W$ drizzled images (340 s each)\footnote{GO-10335, PI: H. Ford}. 
These images were pre-reduced using the HST pipeline.  
To reduce the ACS data, individual PSF were modeled for both chips using DAOPHOT/ALLFRAME 
programs, and the individual catalogs were calibrated following the Sirianni et al.\ (2005) 
prescriptions.

To compare the observations with the models we computed evolutionary tracks with an updated version 
of the FRANEC evolutionary code (Chieffi \& Straniero 1989; Degl'Innocenti et al. 2008). 
For these models the OPAL 2006 equation of state (EOS) was adopted (Iglesias \& Rogers 1996), 
together with radiative opacity tables by the Livermore group 
(Iglesias \& Rogers 1996)\footnote{http://opalopacity.llnl.gov/opal.html} 
for temperatures higher than 12,000$^{o}$K; in this way the EOS
and the opacity are fully consistent. The conductive opacities are
from Shternin \& Yakovlev (2006, see also Potekhin 1999) while the atmospheric opacities
are from Ferguson et al.\ (2005). All the opacity tables were calculated for
the Asplund, Grevesse \& Sauval (2005, hereinafter AG05) solar mixture. The nuclear 
reaction rates, are from the NACRE compilation (Angulo et al.\ 1999).  
It is worth mentioning that the use of the new measurement of the 
$^{14}$N(p,$\gamma$)$^{15}$O capture cross section (Formicola et al.\ 2004) 
would imply a systematic increase of $\sim$1~Gyr in the estimate of the GC 
absolute age (Imbriani et al.\ 2004).    
Our models include atomic
He and metal diffusion, with diffusion coefficients given by Thoul, Bahcall \& Loeb (1994).
The reader interested in a detailed discussion of the uncertainties affecting 
the diffusion coefficients is referred to Bahcall, Pinsonneautl \& Wasserburg (1995), 
Castellani et al.\ (1997) and to Guzik, Watson \& Cox \ (2005).    
To model external convection we adopted, as usual, the mixing length formalism
(Bohm-Vitense 1958). The mixing length parameter, $\alpha$, governing the efficiency
of convection, was set at $\alpha$=2.0.

The metallicity adopted for the models is directly related to the observed
[Fe/H]  (when the  AG05 solar mixture is assumed)
\begin{equation}
\label{eq:zeta}
 Z = \frac{1-Y_P}{1+\frac{\Delta Y}{\Delta Z}+ \frac{1}{(Z/X)_\odot} \times 10^{-[Fe/H]}}. \quad 
\end{equation}

and $\alpha$-element abundances according to formula originally suggested by Salaris et al.\ (1993).    

We adopted the value [Fe/H]=--2.32, given by spectroscopic measurements ([Fe/H]=--2.38) 
by Kraft \& Ivans (2003,2004), but rescaled to the AG05 solar iron abundance. 
The adopted $\alpha$-element  enhancement is  [$\alpha$/Fe]=+0.3, while 
for the primordial helium content, we adopted the recent cosmological value 
$Y_p$=0.248 (Peimbert et al.\ 2007; Izotov, Thuan \& Stasi{\'n}ska \ 2007). 
Note that the change from the old Grevesse \& Sauval (1999) to the new AG05 solar 
mixture causes at fixed iron and $\alpha$-element abundances a decrease from 
$Z$=0.00014 to $Z$=0.00010. The difference is mainly caused by the decrease in CNO solar
abundances. The interested reader is referred to the recent investigation by 
Caffau et al.\ (2010) for a new and independent measurements of solar CNO abundances.
The cluster isochrones were transformed into the observational plane using 
the bolometric corrections (BCs) and the color-temperature relations (CTRs) 
provided by Brott \&  Hauschild (2005), while for the zero-age horizontal-branch (ZAHB) 
models we used the BCs and CTRs provided by Castelli \& Kurucz (2003). 
In the following, the evolutionary models constructed assuming the above chemical 
abundances are called ``canonical models''.

To account for C-N-O-Na anticorrelations, we used the same mixture adopted in 
the literature (Salaris et al.\ 2006; Cassisi et al.\ 2008; Pietrinferni et al.\ 2009) 
which is based on a mean value of the observed anticorrelations provided by 
Carretta et al.\ (2005). The changes in elemental abundance relative to the
canonical $\alpha$-enhanced models are the following:
N increased by 1.8 dex, C decreased by 0.6 dex, Na increased by 0.8 dex,
and O decreased by 0.8 dex. These changes give an enhancement of 
$\approx$ a factor of three (+0.5 dex) in the [C+N+O/Fe] abundance. 
We did not include the Mg--Al anticorrelation because it minimally affects the 
evolutionary properties (Salaris et al.\ 2006). The increase in the total CNO abundance 
causes, at fixed iron abundance, an increase in the total metallicity from $Z$=0.00010 
to $Z$=0.00023. This increase in metallicity makes the MSTO fainter and cooler as 
originally suggested by Bazzano et al.\ (1982) and by VandenBerg (1985).
To constrain the impact of the different chemical mixtures, the CNO-enhanced models 
were constructed both at fixed total metallicity  ([M/H]=--2.10) and at fixed iron 
content ([Fe/H]=--2.32). 
Note that the net effect of element diffusion in $\alpha$-enhanced models is to decrease 
the surface CNO abundances from [C/Fe]=0.00, [N/Fe]=0.00, [O/Fe]=0.30 at the zero age main 
sequence to [C/Fe]=-0.06, [N/Fe]=-0.04, [O/Fe]=0.27 dex at the MSTO. The $\alpha$ and 
CNO-enhanced models show a similar decrese, and indeed the surface CNO abundances change 
from [C/Fe]=-0.60, [N/Fe]=1.80, [O/Fe]=-0.5 at the zero age main sequence to [C/Fe]=-0.65, 
[N/Fe]=1.76, [O/Fe]=-0.53 dex at the MSTO. 

Transforming these evolutionary models into the observational 
plane requires a set of BCs and CTRs computed for the same mixtures, but they are 
not available yet. However, Cassisi et al.\ (2008) and Pietrinferni et al.\ (2009) 
found that BCs and CTRs computed assuming simple $\alpha$-enhanced mixtures mimic the same 
behaviour. Moreover, we found that BCs and CTRs hardly depend, at fixed total 
metallicity, on changes in the mixture.  Accordingly, to transform the CNO-enhanced 
models constructed at fixed iron abundance, we adopted [Fe/H]=--2.32 and 
[$\alpha$/Fe]=+0.72 to obtain the same total metallicity.

%%%%%%%%%%%%%%%%%%%%%%%%%%%%%%%%%%%%%%%%%%%%%%%%%%%%%%%%%%%%%%%%%%%%%%%%%%%%%%%%%%%%%%%%%%%
\section{Comparison between theory and observations}

To compare theory and observations we assumed a true distance modulus $\mu=14.74$ mag, 
in good agreement with the distance estimated by Kraft \& Ivans (2003, $\mu$=14.75)
and a cluster reddening of $E(\bmv)=0.025\pm0.010$ (Zinn 1985; Schlegel, Finkbeiner, \& Davis \ 1989; 
Gratton et al.\ 1997; Kraft \& Ivans 2003; Carretta et al.\ 2000). Moreover, we adopted 
a total to selective absorption 
ratio of $R_V$=3.10 and the empirical reddening laws provided by Cardelli, Clayton \& Mathis \ (1989). 
In particular, for the Sloan bands available at CFHT, we computed the following ratios: 
$A_{g'}/A_V$=$1.21$, $A_{r'}/A_V$=0.87, $A_{i'}/A_V$ =0.66 and $A_{z'}/A_V$=0.48.  
For the Johnson-Cousins filters we adopted $A_B/A_V$=1.32 and $A_I/A_V$=0.59, while for 
the ACS filters $A_{F475W}/A_V$=1.20 and $A_{F814W}/A_V$=0.55, respectively.

The top panels of Fig.~3 show the comparison between data collected in different 
photometric bands and $\alpha$-enhanced isochrones at fixed total metallicity 
([M/H]=--2.10 dex) and two cluster ages 10 (red line) and 12 (green line)~Gyr.  
The CFHT data (the first three panels from left to right) were selected according  
to photometric error ($\sigma$(color)$\le$ 0.04), sharpness ($\textit{abs(sha)}\le 1$) 
and radial distance ($60\farcs \le r \le 400 \farcs$).  
The same selection was adopted for the Johnson data (fourth panel), while the 
ACS data (fifth panel) were only selected on the basis of the photometric 
error.
To validate the adopted values of the true distance modulus and the cluster reddening we 
also plotted the predicted ZAHB for the same chemical 
composition (blue line). We found that using the same distance and the same reddening 
the predicted ``canonical'' ZAHB agrees quite well with observations in the five different CMDs. 
Moreover, the comparison between theory and observations gives a cluster age
of $11\pm1.5$~Gyr. The error budget is mainly driven by observational uncertainties affecting the metallicity 
measurements, and theoretical uncertainties in the input physics of the evolutionary models and transformations.     

To constrain the impact of the CNO abundance on the age of M92, we performed the same 
comparison at fixed total metallicity, but using the CNO-enhanced models.
The middle panels of Fig.~3 show the same data as plotted in the top 
panel, but the cluster isochrones are based on evolutionary models that are enhanced 
in both $\alpha$ elements and CNO. The same enriched composition 
was also adopted to compute the ZAHB and we found no significant differences compared to the 
canonical models. Therefore, we assumed the same true distance modulus and cluster reddening 
as for the canonical predictions. The comparison between 
theory and observations indicates that $\alpha$ and CNO enriched isochrones provide, 
at fixed total metallicity, the same cluster age ($11\pm1.5$~Gyr) as the canonical 
$\alpha$-enhanced models.  Data plotted in the bottom panels show the comparison
between theory and observations for $\alpha$ and CNO enhancements at fixed iron
abundance ([Fe/H]=--2.32). Note that the total metallicity of the isochrones and 
of the ZAHB is now increased to [M/H]=--1.75. The cluster age we found is, for the same true 
distance and cluster reddening, minimally younger ($10\pm1.5$~Gyr), but still agrees quite 
well ($\pm 1~\sigma$) with the above estimates. It is worth noticing that canonical and 
CNO-enhanced isochrones show very similar morphologies in optical CMDs, 
as originally suggested by Salaris et al.\ (2006).  
This indicates that broad-band optical photometry like that investigated here cannot be safely 
adopted to constrain the occurrence of CNO-enhanced subpopulations in GCs.

%%%%%%%%%%%%%%%%%%%%%%%%%%%%%%%%%%%%%%%%%%%%%%%%%%%%%%%%%%%%%%%%%%%%%%%%%%%%%%%%%%%%%%%%%%%%

To further constrain the impact of the CNO enhancement on the evolutionary properties of metal-poor 
GCs we also compared the observed star count ratios with the evolutionary lifetime ratios. The 
comparison between theory and observations was performed using evolved (RGB, HB) and 
MS stars. The MS stars were selected in a magnitude interval of 0.25 mag across the 
MSTO ($M_{i'}=18.8$), while the RGB stars were selected in the magnitude interval 
15.0 $\le$ $i'$ $\le$ 17.2 (see the left panel of Fig.~4). 
To provide robust star counts over the entire body of the cluster we used 
ACS/HRC data for the regions across the very center of the cluster ($\le 35\farcs$), the 
ACS/WFC data up to radial distances of $75\farcs$ and the CFHT data for the remaining 
regions. 
To homogenize the star counts, the ACS/HRC and the ACS/WFC data collected in the 
$F814W$-band were transformed into the $i'$-band, while those ones collected in the
$F475W$ and in the $F555W$-band were transformed into the $g'$ and $r'$-band.
The accuracy of the transformations is better than 0.02 mag.    
To estimate the completeness of the CFHT data in the regions outside the internal ACS/WFC 
pointing, we adopted the ACS/WFC data of the external pointing. We found that the 
completeness for $i'$$\le$ 22 mag is $\sim$54\%  for 75\farcs $\le$ r $\le$ 150\farcs, 
$\sim$82\%  for 150\farcs $\le$ r $\le$ 200\farcs, and complete for larger distances.   
To investigate possible radial trends (Castellani et al.\ 2007; Sandquist \& Martel 2007) 
the cluster was divided into eight annuli up to r=400$\farcs$. Each annulus includes 
the same number of stars ($\sim$ 12,200).

The right panel of Fig.~4 shows the star count ratios HB/RGB (top), RGB/MS (middle) and 
HB/MS (bottom) as a function of the radial distance. Data plotted in this figure show the star counts 
are, within the errors, constant across the cluster. To estimate the lifetime ratios 
of the same evolutionary phases we adopted for the $\alpha$-enhanced models the evolutionary 
track of a stellar structure with $M(TO)/M_\odot$=0.78, [M/H]=--2.10. Note that this is the 
TO mass of the 11 Gyr cluster isochrone. For the $\alpha$- and CNO-enhanced models with the 
same total metallicity we adopted the same stellar mass, while for those with the same 
iron abundance ([Fe/H]=--2.32) we adopted a stellar mass of $M(TO)/M_\odot$=0.80. This is 
the TO mass of the 10 Gyr cluster isochrone. The HB evolutionary lifetime was estimated 
using a structure with stellar mass of $M/M_\odot$=0.70. This mass value relies on the 
mean mass for HB stars found by Cassisi et a.\ (2001) and by Cho \& Lee \ (2007) using 
synthetic HB models. Note that a change of 0.05 $M_\odot$ has a minimal impact on the 
HB lifetime (3-4\%).             
By using the same true distance modulus and 
cluster reddening adopted to estimate the cluster age, we found that observed and predicted 
ratios agree quite well (1$\sigma$, see Table 1). It is worth mentioning that the stated uncertainties 
in the star counts include Poisson uncertainties and completeness uncertainties. For the 
predicted ratios we assumed an uncertainty of 10\% for each of the different evolutionary phases
(Castellani et al.\ 2007, and references therein).

The anonymous referee suggested that we specify the impact of
the different chemical mixtures on both the opacities and the
burning processes. To disentangle the effects, we performed a
series of numerical experiments following the approach recently
adopted by Pietrinferni et al.\ (2009) for a chemical composition
that is at the metal-rich end of GCs, namely [Fe/H]=--0.7 dex, by
Ventura et al.\ (2009) for the metal-intermediate
([Fe/H]$\sim$~--1.2 dex) GC NGC~1851 and by Di~Criscienzo et al.\
(2010) for the metal-poor ([Fe/H]$\sim$~--2 dex) GC NGC~6397. We
performed these experiments at fixed iron content to reveal the
differential effects produced by either $\alpha$-enhanced or
$\alpha$- plus CNO-enhanced mixtures.  The green solid and the
blue dashed lines plotted in Fig.~5 show two evolutionary tracks
constructed at fixed stellar mass (M=0.80 M$_\odot$), iron content
([Fe/H]=--2.32) and helium content ($Y$=0.248), but the former
assumes a canonical $\alpha$-enhanced mixture ([M/H]=--2.10,
$Z$=0.00010), while the latter assumes the $\alpha$- and
CNO-enhanced mixture ([M/H]=--1.75, $Z$=0.00023).  As expected,
the difference is minimal along the MS, but becomes relevant
between the TO and the base of the RGB. During these evolutionary
phases the track with the $\alpha$- and CNO-enhanced mixture is,
at fixed effective temperature, systematically fainter. The MSTO
is fainter and cooler.  The RGB bump, i.e., the evolutionary
phases during which the H-burning shell encounters the chemical
discontinuity in the envelope left over by the first dredge-up is
slightly fainter in the $\alpha$- and CNO-enhanced track than in
the $\alpha$-enhanced track. The difference is caused by the fact
that the convective envelope during the earlier RGB phases deepens
more in the former than in the latter case (Salaris et al.\
2006; Pietrinferni et al.\ 2009; Di Cecco et al.\ 2010),  because
the increase in the total metallicity causes an increase in the
opacity.  

To further investigate the specific impact of $\alpha$- and
CNO-enhanced mixture on the opacity we also constructed an
evolutionary track in which the nuclear burning processes use an
$\alpha$-enhanced mixture, while the opacity is based on an
$\alpha$ and CNO-enhanced mixture. The red dashed-dotted line
plotted in Fig.~5 shows that this track is, at fixed effective
temperature, slightly fainter than the track using an
$\alpha$-enhanced mixture (green solid line) both in the nuclear
burning and in the opacity.  As a consequence, the MSTO of the
former track is marginally cooler, while the RGB bump is minimally
fainter than in the latter one.  

To study the impact of $\alpha$- and CNO-enhanced mixture on the
nuclear network we also constructed an evolutionary track in which
the nuclear burning processes use an $\alpha$- and CNO-enhanced
mixture, while the opacity is based on an $\alpha$-enhanced
mixture. The pink dotted line plotted in Fig.~5 shows that this
track is, at fixed effective temperature, minimally brighter than
the track with an $\alpha$- and CNO-enhanced mixture  both in the
nuclear burning and in the opacity. The same outcome applies to
the MSTO and to the RGB bump.  

These findings indicate that the changes in the luminosity and in
the effective temperature of the MSTO, when switching from an
$\alpha$-enhanced to an $\alpha$- and CNO-enhanced mixture, are
caused by changes both in the nuclear burning and in the opacity.
On the other hand, the change in the luminosity of the SGB is
caused primarily by the change in the nuclear burning processes.
The same outcome applies for the decrease in the luminosity of the
RGB bump. These results support the dependences found by
Pietrinferni et al.\ (2009) for more metal-rich stellar
structures.

%%%%%%%%%%%%%%%%%%%%%%%%%%%%%%%%%%%%%%%%%%%%%%%%%%%%%%%%%%%%%%%%%%%%%%%%%%%%%%%%%%%%%%%%%%%%%%%%%%%%
\section{Conclusions}

We present different optical data sets for M92 collected in three different photometric systems 
(SDSS,  $g',r',i',z'$; Johnson-Cousins, $BVI$; ACS Vegamag, $F475W$, $F555W$, $F814W$) 
and with ground-based and space (HST) telescopes. Special attention was given to the precision of the 
photometric zero-points. By using the same true distance modulus and cluster reddening 
our canonical $\alpha$-enhanced isochrones constructed assuming [Fe/H]=--2.32, [$\alpha$/Fe]=0.3 
and $Y$=0.248, account for the observed features in five different CMDs. We found a cluster age of 
 11$\pm$1.5 Gyr, supporting previous results based on cluster isochrones and luminosity functions. 
The same outcome applies to the comparison between the HB stars and the predicted ZAHB.        
We also investigated the impact of a CNO enriched chemical composition and we found that 
$\alpha$- and CNO- enhanced isochrones at fixed total metallicity ([M/H]=--2.10) provide, 
within the errors, the same cluster age. Moreover, $\alpha$- and CNO- enhanced isochrones 
at fixed iron abundance ([Fe/H]=--2.32, [M/H]=--1.75)  give a cluster age that is minimally 
younger (10$\pm$1.5 Gyr).        

We also investigated the star count ratios for evolved (RGB,HB) and MSTO stars.
We found that they do not show any radial trend when moving from the very center to the 
outermost cluster regions. Moreover and even more importantly, star count ratios agree 
quite well (within 1$\sigma$) with the lifetime ratios of the same evolutionary phases. 

The above results indicate that the occurrence of CNO enriched subpopulations has a 
minimal impact on the cluster age in the metal-poor domain. The same outcome applies 
to star count ratios and evolutionary lifetimes. These findings appear quite robust, 
since they rely on different photometric data sets covering the entire body of the 
cluster and on the same evolutionary framework. 

We also note that isochrones including atomic He and metal diffusion give cluster 
ages that are $\approx$1 Gyr younger than canonical isochrones (Castellani et al.\ 1997). 
This means that current findings support previous theoretical predictions for typical 
GCs by Salaris et al.\ (2006) and recent age estimates for metal-poor GCs provided by 
Marin-Franch et al.\ (2009). 
Finally, it is worth emphasizing that current age estimates agree quite well with
the cluster age provided by VandenBerg et al.\ (2002) using an independent but similar 
theoretical framework. The difference is larger than one $\sigma$ only for the cluster 
age based on the CNO-enhanced models computed at fixed iron content 
(10$\pm$1.5 vs 13.5$\pm$1.5 Gyr). This difference can be explained if we account for 
the mild change in the shape of the SGB region of these isochrones when compared with 
the canonical ones.
   
%%%%%%%%%%%%%%%%%%%%%%%%%%%%%%%%%%%%%%%%%%%%%%%%%%%%%%%%%%%%%%%%%%%%%%%%%%%%%%%%%%%%%
% 			fig 3
%%%%%%%%%%%%%%%%%%%%%%%%%%%%%%%%%%%%%%%%%%%%%%%%%%%%%%%%%%%%%%%%%%%%%%%%%%%%%%%%%%%%%
 
\begin{figure*}[t]
\begin{center}
\label{fig3}
\includegraphics[height=24cm,width=18cm]{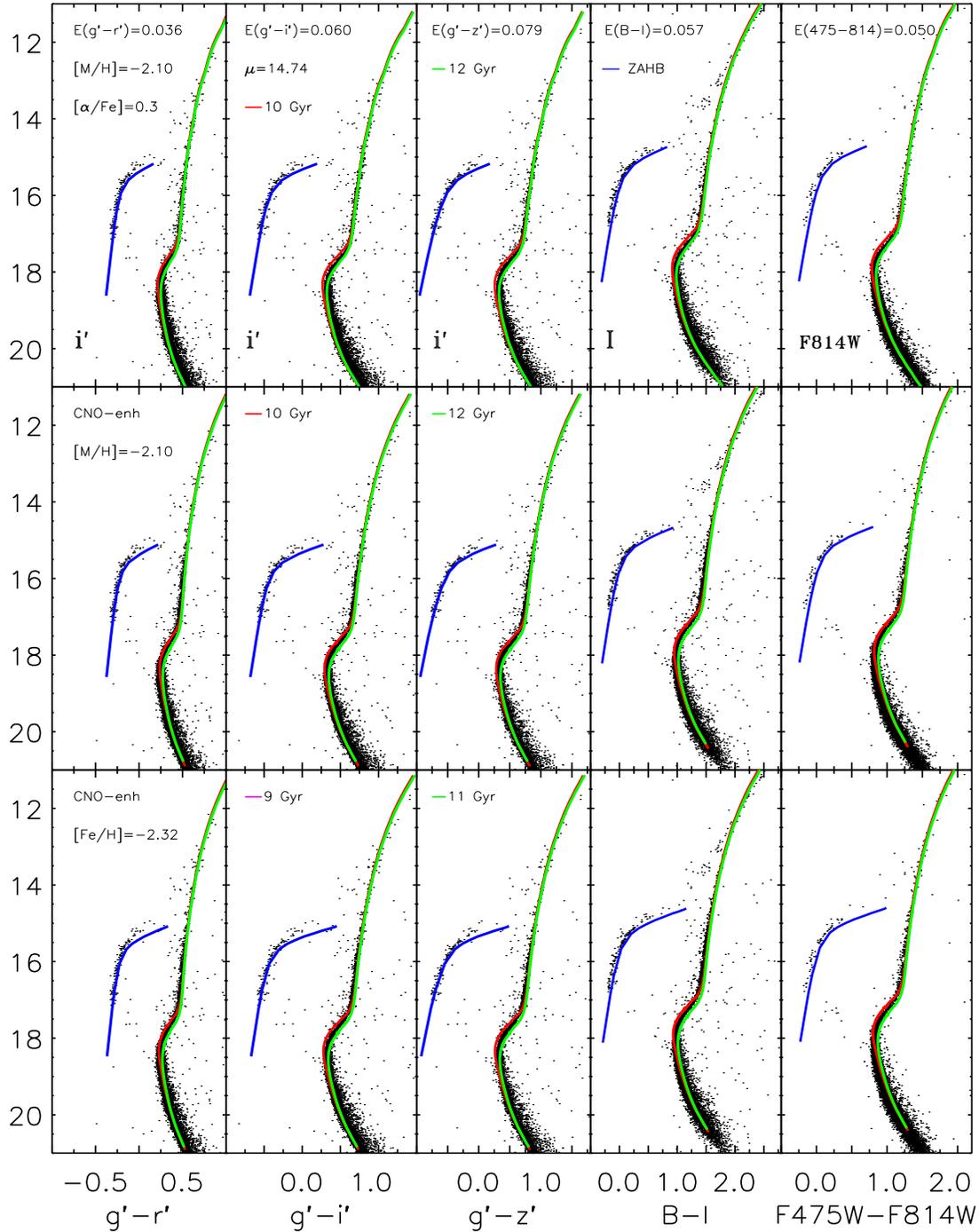}
\vspace*{-3.15truecm}
\caption{Top -- from left to right the first three panels show the CMDs
($i'$,$g'-r'$; $i'$,$g'-i'$; $i'$,$g'-z'$) based on CFHT data. The fourth and
the fifth panel show the Johnson-Cousins (I,$\bmi$) and the ACS (F814W, F475W-F814W)
CMDs. The red and the green lines display 10 and 12 Gyr $\alpha$-enhanced 
isochrones at fixed chemical composition (see labeled values). The blue lines 
show the predicted ZAHB for the same chemical composition. The adopted true 
distance modulus and cluster reddenings are also labeled.   
Middle -- same as the top, but the ZAHB and the isochrones are based on 
evolutionary models constructed assuming an $\alpha$- and CNO-enhanced mixture. 
Predictions plotted in these panels have the same total metallicity ([M/H]=--2.10) 
of the $\alpha$-enhanced predictions (top).
Bottom -- same as top, but for 9 (red line) and 11 (green line) Gyr isochrones. 
The ZAHB and the isochrones are based on evolutionary models constructed assuming an 
$\alpha$- and CNO-enhanced mixture. Predictions plotted in these panels have the 
same iron abundance ([Fe/H]=--2.32) of the $\alpha$-enhanced predictions (top). }
\end{center}
\end{figure*}

%%%%%%%%%%%%%%%%%%%%%%%%%%%%%%%%%%%%%%%%%%%%%%%%%%%%%%%%%%%%%%%%%%%%%%%%%%%%%%%%%%%%%
% 			fig 4
%%%%%%%%%%%%%%%%%%%%%%%%%%%%%%%%%%%%%%%%%%%%%%%%%%%%%%%%%%%%%%%%%%%%%%%%%%%%%%%%%%%%%
\begin{figure*}
\begin{center}
\label{fig4}
\includegraphics[height=11cm,width=16cm]{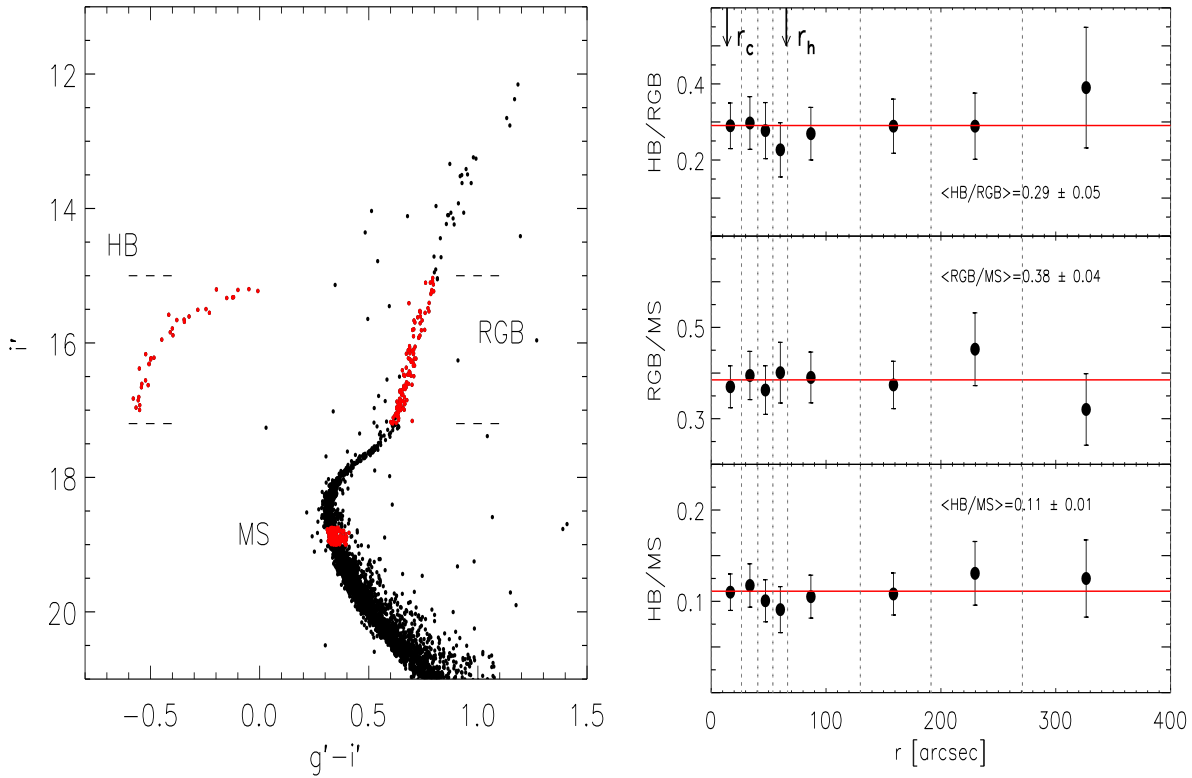}
\caption{\footnotesize{Left -- $i'$,$g'-i'$ CMD based on CFHT data. The red dots mark the 
selected HB, RGB and MS subsamples. Right -- Star count ratios HB/RGB (top), RGB/MS (middle) and 
HB/MS (bottom) as a function of the radial distance (r$\le 400\farcs$). The mean values and the 
standard deviations are labeled. Vertical dotted lines display the radial extent of the different annuli adopted 
to estimate the star counts. The error bars account for Poisson and completeness uncertainties. 
The two vertical arrows mark the core radius ($r_c \sim 14\farcs$) and the half-mass radius 
($r_h \sim 65\farcs$).}}
\end{center}
\end{figure*}

%%%%%%%%%%%%%%%%%%%%%%%%%%%%%%%%%%%%%%%%%%%%%%%%%%%%%%%%%%%%%%%%%%%%%%%%%%%%%%%%%%%%%
% 			fig 5
%%%%%%%%%%%%%%%%%%%%%%%%%%%%%%%%%%%%%%%%%%%%%%%%%%%%%%%%%%%%%%%%%%%%%%%%%%%%%%%%%%%%%
\begin{figure*}
\begin{center}
\label{fig4}
\includegraphics[height=18cm,width=16cm]{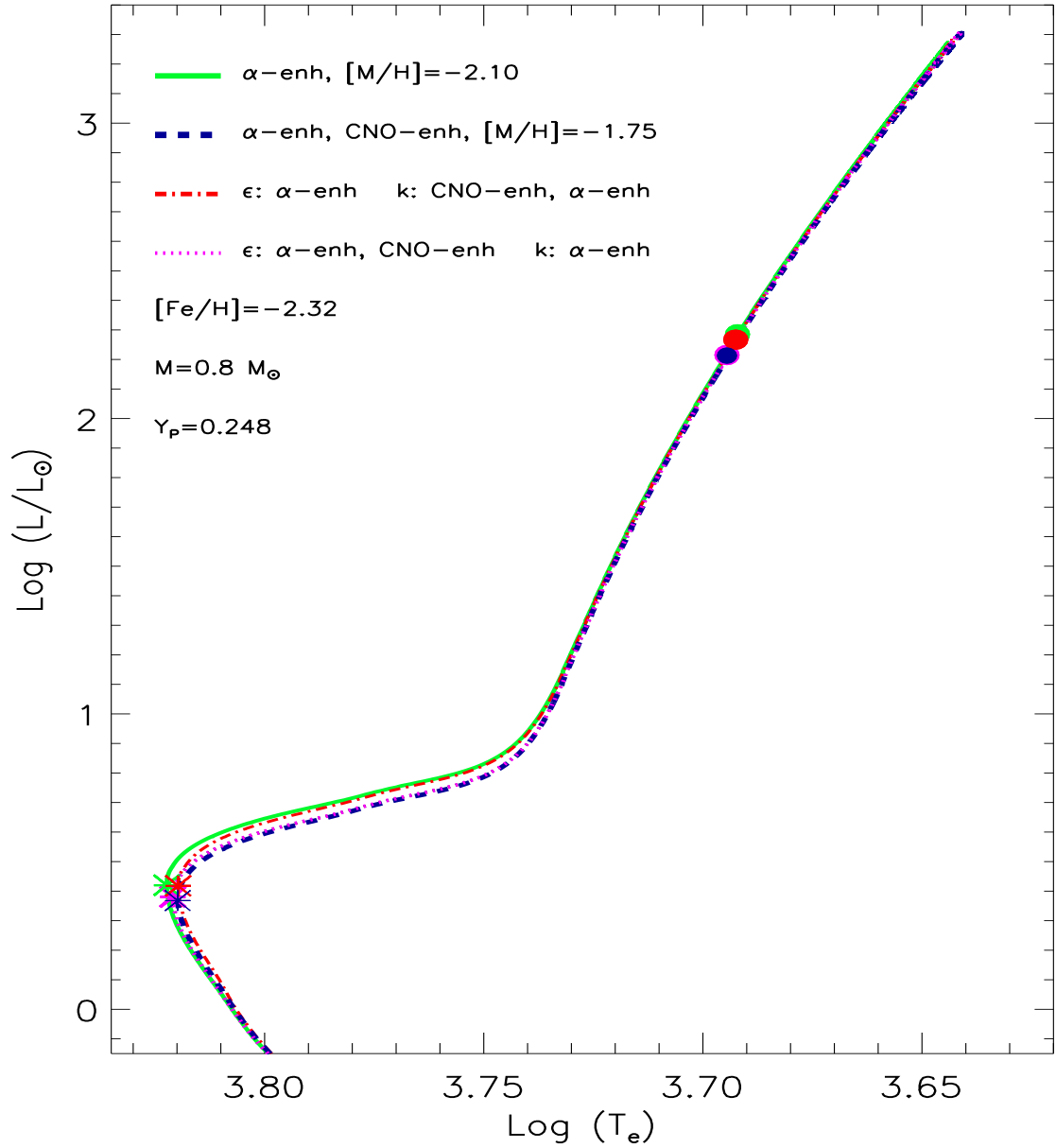}
\caption{\footnotesize{Hertzsprung-Russell diagram of low-mass evolutionary tracks 
constructed at fixed stellar mass, helium and iron content (see labeled values), 
but different chemical mixtures. The green solid and the blue dashed line show 
two tracks constructed assuming a canonical $\alpha$-enhanced and an $\alpha$ and 
CNO-enhanced chemical mixture. The dashed-dotted line shows a track in which the 
nuclear burning uses an $\alpha$-enhanced mixture, while the opacity an $\alpha$ and
CNO-enhanced mixture. The dotted track shows a track in which the nuclear burning uses 
an $\alpha$ and CNO-enhanced mixture, while the opacity an $\alpha$-enhanced mixture.       
The asterisks and circles mark the position of the MSTO and of the RGB bump.
}} 
\end{center}
\end{figure*}

%%%%%%%%%%%%%%%%%%%%%%%%%%%%%%%%%%%%%%%%%%%%%%%%%%%%%%%%%%%%%%%%%%%%%%%%%%%%%%%%%%%%%
% 			tab
%%%%%%%%%%%%%%%%%%%%%%%%%%%%%%%%%%%%%%%%%%%%%%%%%%%%%%%%%%%%%%%%%%%%%%%%%%%%%%%%%%%%%
\begin{deluxetable}{lccc}
\scriptsize
\tablewidth{0pt}                       
\tablecaption{Star count ratios and evolutionary lifetime ratios.}
\tablehead{
\colhead{Ratios}&
%\colhead{l\tablenotemark{b}}&
\colhead{HB/RGB}&
\colhead{RGB/MS}&
\colhead{HB/MS}
}
\startdata
Empirical$^a$ & $ 0.29\pm0.05 $ & $  0.38\pm0.04 $& $ 0.11\pm0.01 $\\
Theory$^b$    & $ 0.31\pm0.04 $ & $  0.33\pm0.05 $& $ 0.10\pm0.01 $\\
Theory$^c$    & $ 0.29\pm0.04 $ & $  0.33\pm0.05 $& $ 0.10\pm0.01 $\\
Theory$^d$    & $ 0.32\pm0.04 $ & $  0.30\pm0.04 $& $ 0.09\pm0.01 $\\ 
\enddata
\tablenotetext{a}{Mean star count ratios from the very center up to $\sim$7\farcm}  
\tablenotetext{b}{Lifetime ratios based on an evolutionary model constructed 
assuming $M(TO)/M_{\odot}$=0.78, and an $\alpha$-enhanced mixture ([Fe/H]=--2.32, 
[$\alpha$/Fe]=0.3).}  
\tablenotetext{c}{Lifetime ratios based on an evolutionary model constructed 
assuming $M(TO)/M_{\odot}$=0.78, and an $\alpha$- and CNO-enhanced mixture, but 
the same total metallicity  ([M/H]=--2.10) as the $\alpha$-enhanced model.}  
\tablenotetext{d}{Lifetime ratios based on an evolutionary model constructed 
assuming $M(TO)/M_{\odot}$=0.80, and an $\alpha$- and CNO-enhanced mixture, but
the same iron abundance ([Fe/H]=--2.32) as the $\alpha$-enhanced model.}
\end{deluxetable}

\acknowledgments
It is a real pleasure to thank an anonymous referee for his/her positive
comments on the results of this investigation and for his/her suggestion.
We also thank S. Cassisi and A. Pietrinferni  for several useful discussions 
concerning  low-mass stars and chemical mixtures. 
This project was partially supported by the  grant Monte dei Paschi di
Siena (P.I.: S. Degl'Innocenti) and  PRIN-MIUR2007 (P.I.: G. Piotto).

%--------------------------------------------------------------------------------------

%
%


\begin{thebibliography}{}

\bibitem[Angulo et al. (1999)]{angu} Angulo C. et al., 1999, Nucl. Phys., 656, 3

\bibitem[Asplund \& Grevesse (2005)]{aspl} Asplund, M., Grevesse, N., \&
 Sauval, A.J. 2005, in Cosmic Abundances as Records of Stellar Evolution and
 Nucleosynthesis, eds. F.N. Bash, \& T.J. Barnes, ASP Conf. Series, 336, 25 [AG05]

\bibitem[bahcall (1995)]{ba95}Bahcall, J. N., Pinsonneault, M. H., \& Wasserburg, G. J. 1995, RevModPhys, 67, 781 
	
\bibitem[bazzano et al. (1982)]{baz}Bazzano, A., Caputo, F., Sestili, M., \& Castellani, V. 1982,  A\&A, 111, 312

%\bibitem[Bedin et al. (2004)]{bed04} Bedin, L. R., Piotto, G., Anderson, J., Cassisi, S., King, I. R., Momany, Y., \& Carraro,
%G. 2004, ApJ, 605, L125

\bibitem[belman et al.(2001)]{bel}Bellman, S., Briley, M. M., Smith, G. H., \& Claver, C. F. 2001, PASP, 
113, 326

%\bibitem[bhom (1976)]{boh}Bohm-Vitense, E.; \& Nelson, G. D. 1976, ApJ, 210, 741
	
\bibitem[bhom (1958)]{boh58} Bohm-Vitense, E. 1958, Zeitschrift f\"ur Astrophysik, 46, 108
	
\bibitem[bono (2001)]{bon1} Bono, G., Caputo, F., Castellani, V., Marconi, M., \& Storm, J., 2001, MNRAS, 326, 1183

\bibitem[bono (2008)]{bon2} Bono, G. et al. 2008, ApJ, 686, 87   

\bibitem[Brott et al.(2005)]{brott} Brott, I.,\& Hauschildt, P. H. 2005, ESASP, 576, 565

\bibitem[Caffau et al (2019)]{caffau2010} Caffau, E. et al. 2010, A\&A, 514, 92

\bibitem[cannon (1998)]{can1} Cannon, R. D., Croke, B. F. W., Bell, R. A., Hesser, J. E., \& Stathakis, R. A. 1998,  MNRAS, 298, 601
	
\bibitem[carbon (1982)]{carbon}Carbon, D. F., Romanishin, W., Langer, G. E., Butler, D., Kemper, E., Trefzger, C. F., Kraft,
R. P., \& Suntzeff, N. B. 1982,  ApJS, 49, 207

\bibitem[card (1989)]{card} Cardelli, J. A., Clayton, G. C., \& Mathis, J. S. 1989, IAUS, 135, 5

\bibitem[carretta et al (2009)]{c09} Carretta, E., Bragaglia, A., Gratton, R., D'Orazi, V., \& Lucatello, S.  2009, A\&A, 508, 695

\bibitem[Carretta et al. (2007)]{c07} Carretta, E., Bragaglia, A., Gratton, R. G., Lucatello, S., \& Momany, Y.  2007, A\&A
, 464, 927

\bibitem[Carretta et al.(2000)]{car00} Carretta, E., Gratton, R. G., Clementini, G., \& Fusi Pecci, F. 2000, ApJ, 533, 215

 \bibitem[Carretta et al. (2005)]{c05} Carretta, E., Gratton, R. G., Lucatello, S., Bragaglia, A., \&  Bonifacio, P. 2005, A\&A, 433, 597
 
\bibitem[cassi (2004)]{cas04} Cassisi, S., Castellani, M., Caputo, F.,\& Castellani, V., 2004, A\&A, 426, 641
 
 \bibitem[Cassisi et al. (2001)]{cas01} Cassisi, S., Castellani, V.,  Degl'Innocenti, S., Piotto, G., \& Salaris, M. 2001, A\&A, 366, 578
 	
\bibitem[cassi (2008)]{cassi2008} Cassisi, S., Salaris, M., Pietrinferni, A., Piotto G. , Milone, A. P., Bedin, L. R., \& Anderson, J.  2008, ApJ, 672, L115

\bibitem[castellani (2007)]{cast07}Castellani, V. et al. 2007, ApJ, 663, 1021

\bibitem[castellani (1997)]{c1997}Castellani, V., Ciacio, F., degl'Innocenti, S., \& Fiorentini, G. 1997, A\&A, 322, 801
	
\bibitem[Castelli et al.(2003)]{castelli} Castelli, F.,\& Kurucz, R. L. 2003, IAUS, 210, A20

\bibitem[cate (2004)]{cat04}Catelan, M. 2004,  ASPC, 310, 113

%\bibitem[Chaboyer (1998)]{cha98}Chaboyer, B. 1998  

\bibitem[Chieffi \& Straniero (1989)]{chieffi} Chieffi A., \& Straniero O. 1989, ApJS, 71, 47

 \bibitem[Cho et al.(2007)]{cho} Cho, D. H., \& Lee, S. G. 2007, AJ, 133, 2163 
 
 \bibitem[clem (2007)]{clem07}Clem, J. L., VandenBerg, D. A., \& Stetson, P. B. 2007,  AJ, 134, 1890

\bibitem[Cohen (1978)]{coh78}Cohen, J. G. 1978, ApJ, 223, 487

\bibitem[dancal (2008)]{dant} D'Antona, F.,\& Caloi, V. 2008, MNRAS, 390, 693

\bibitem[decre (2007)]{decre}Decressin, T., Meynet, G., Charbonnel, C., Prantzos, N., \& Ekstr\"om, S. 2007, A\&A, 464, 1029

\bibitem[Degl'Innocenti et al. (2008)]{deg08} Degl'Innocenti S., Prada Moroni
P. G., Marconi M., \& Ruoppo A., 2008, Astrophysics and Space Science, 316, 215 

\bibitem[delprincipe (2006)]{delp2}Del Principe, M. et al. 2006,  ApJ, 652, 362

\bibitem[delprincipe (2005)]{delp1}Del Principe, M., Piersimoni, A. M., Bono, G., Di Paola, A., Dolci, M., \& Marconi, M. 2005,  AJ, 129, 2714

\bibitem[Denissenkov \& Weiss (2004)]{den} Denissenkov, P. A., \& Weiss, A. 2004, ApJ, 603, 119

\bibitem[Di Cecco et al. (2010)]{dicecco2010} Di Cecco, A. et al. 2010, ApJ, 712, 527

\bibitem[Di Criscienzo et al (2010)]{dicre2010} Di Criscienzo, M., D'Antona, F., \& Ventura, P. 2010, A\&A, 511, 70

\bibitem[Ferguson et al. (2005)]{ferg} Ferguson J.W., Alexander, D. R., Allard, F., Barman, T., Bodnarik, J. G., Hauschildt,
P. H., Heffner-Wong, A., \& Tamanai, A. 2005, ApJ, 623, 585

\bibitem[Formicola et al. (2004)]{formy}Formicola, A. et al. 2004, Phys. Lett. B, 591, 61

\bibitem[grat (2001)]{grat01}Gratton, R. G. et al. 2001, A\&A, 369, 87

\bibitem[gra (2003)]{gra2003} Gratton, R. G., Bragaglia, A., Carretta, E., Clementini, G. , Desidera, S., Grundahl, F.;, \& Lucatello, S. 2003, A\&A, 408, 529

\bibitem[gratt (1997)]{grat97}Gratton, R. G., Fusi Pecci, F., Carretta, E., Clementini, G., Corsi, C. E., \& Lattanzi, M. 1997, ApJ, 491, 749
	
\bibitem[Gratton et al. (2004)]{gra04} Gratton, R., Sneden, C., \& Carretta, E. 2004, A\&S, 42, 385   

\bibitem[Grevesse \& Sauval (1999)]{greve99} Grevesse, N., \& Sauval, A. J. 1999, A\&A, 347, 348

\bibitem[gru (2000)]{grun}Grundahl, F., VandenBerg, D. A., Bell, R. A., Andersen, M. I.,\& Stetson, P. B. 2000, AJ, 120, 1884

\bibitem[guzik (2005)]{gu05} Guzik, J. A., Watson, L. S., \& Cox, A. N.	2005, ApJ, 627, 1049    

\bibitem[har (2003)]{harb} Harbeck, D., Grebel, E. K., \& Smith G. H. 2003,  ANS, 324, 78

\bibitem[Harris (1996)]{har96}Harris, W. E. 1996, AJ, 112, 1487

\bibitem[Iglesias \& Rogers (1996)]{ig96} Iglesias, C., \& Rogers, F.J., 1996, ApJ, 464, 943

\bibitem[Imbriani et al. (2004)]{imbri04} Imbriani, G. et al. 2004, A\&A, 420, 625

\bibitem[Izotov et al. (2007)]{Izot} Izotov, Y. I., Thuan, T. X., \& Stasi{\'n}ska, G., 2007, 662, 15

\bibitem[Kraft (1994)]{kraft} Kraft, R. P., 1994, PASP, 106, 553

\bibitem[Kraft \& Ivans (2003)]{kra03} Kraft, R. P., \& Ivans, I.I. 2003, PASP, 115, 143

\bibitem[Kraft \& Ivans (2004)]{kra04} Kraft, R. P., \& Ivans, I.I. 2004, arXiv:astro-ph/0305380v1

\bibitem[Landolt (1992)]{land92} Landolt, A. U. 1992, AJ, 104, 340

\bibitem[langer (1986)]{langer} Langer, G. E., Kraft, R. P., Carbon, D. F., Friel, E., \& Oke, J. B. 1986,  PASP, 98, 473

\bibitem[Leep et al. (1986)]{leep86} Leep, E. M., Wallerstein, G., \& Oke, J. B. 1986, AJ, 91, 1117

\bibitem[maeder (2006)]{maeder} Maeder, A., \& Meynet, G. 2006,  A\&A, 448, L37

\bibitem[magner (2004)]{magnier} Magnier, E. A., \& Cuillandre, J.-C. 2004, PASP, 116, 449

\bibitem[marin (2009)]{marin} Marin-Franch, A. et al. 2009, ApJ, 694, 1498

\bibitem[Norris et al. (1981)]{nor81} Norris, J., Cottrell, P. L., Freeman, K.C., \& Da Costa, G. S. 1981, ApJ, 244,205

\bibitem[Osborn (1971)]{os71} Osborn, W. 1971, The Observatory, 91, 223

\bibitem[pau (2007)]{paust}Paust, N. E. Q., Chaboyer, B., \& Sarajedini, A. 2007, AJ, 133, 2787

\bibitem[Peimbert et al. (2007)]{Peimb} Peimbert, M., Luridiana, V., Peimbert, A., \& Carigi, L. 2007, Astronomical Society of the Pacific Conference Series, 374, 81

\bibitem[Peterson (1980)]{peter} Peterson, R. C.  1980, ApJ, 237, 87
	
 \bibitem[Pietrinferni et al.(2009)]{piet1} Pietrinferni, A., Cassisi, S., Salaris, M., Percival, S., \& Ferguson, J. W. 2009, ApJ, 697, 275

%\bibitem[pila (1988)]{pila} Pilachowski, C.A., 1988,  ApJ, 326, L57

\bibitem[pila (1983)]{pil83}Pilachowski, C. A., Bothun, G. D., Olszewski, E. W., \& Odell, A. 1983, ApJ,
273, 187

\bibitem[pio07 (2007)]{Piotto07}Piotto, G. et al. 2007, ApJ, 661, L53

\bibitem[Potekhin (1999)]{potek} Potekhin A.Y., 1999, A\&A, 351, 787

\bibitem[pra (2006)]{prant} Prantzos, N.,\& Charbonnel, C. 2006,  A\&A, 458, 135

\bibitem[ramirez (2002)]{rami} Ramirez, S. V., \& Cohen, J. G. 2002, AJ, 123, 3277

%\bibitem[Reid (1997)]{Reid97} Reid, I. N., 1997, AJ, 114, 161

\bibitem[ren (1991)]{renzini}Renzini, A. 1991, in Observational Test of Cosmological Inflaction, ed. T. Shanks, A. J. Banday, \& R. S. Ellis (Dordrecht: Kluwer), 131]

\bibitem[Salaris et al. (1993)]{sala93} Salaris, M., Chieffi, A., \& Straniero, O. 1993, ApJ, 414, 580

\bibitem[Salaris \& Weiss (2002)]{ss02} Salaris, M. \& Weiss, A. 2002, A\&A, 388, 492

\bibitem[Salaris et al.(2006)]{sala} Salaris, M., Weiss, A.,  Ferguson, J. W., \& Fusilier, D. J. 2006, ApJ, 645, 1131

\bibitem[sand (2007)]{sand}Sandquist, E. L., \& Martel, A. R. 2007, ApJ, 654L, 65

\bibitem[schlegel (19989]{schle} Schlegel, D. J., Finkbeiner, D. P., \& Davis, M. 1998,  ApJ, 500, 525

\bibitem[Shternin \& Yakovlev(2006)]{shtern} Shternin, P.S. \& Yakovlev, D.G., 2006, PhRvD, 74, (4) 3004

%\bibitem[siri (2003)]{siri}Sirianni, M. et al. 2003,  AAS, 202, 1701

\bibitem[siri (2005)]{siri05} Sirianni, M. et al. 2005, PASP, 117, 1049
	
\bibitem[Smith (1987)]{smi87} Smith, G. H., 1987, PASP, 99, 67

\bibitem[sneden (1991)]{snede}Sneden, C., Kraft, R. P., Prosser, C. F., \& Langer, G. E. 1991,  AJ, 102, 2001

\bibitem[sollima (2006)]{sollima} Sollima, A., Pancino, E., Ferraro, F. R., Bellazzini, M., Straniero, O., \& Pasquini, L. 2005,  ApJ, 634, 332

\bibitem[ste1 (1987)]{stet1} Stetson, P. B. 1987, PASP, 99, 191

\bibitem[ste2 (1994)]{stet2} Stetson, P. B. 1994, PASP, 106, 250

\bibitem[stetson (2000)]{stet3} Stetson, P. B. 2000, PASP, 112, 925

\bibitem[stetson (2005)]{stet05} Stetson, P. B. 2005, PASP, 117, 563

\bibitem[stetson (2003)]{stet03} Stetson, P. B., Bruntt, H., \& Grundahl, F.  2003, PASP, 115, 413

\bibitem[stetson (2004)]{stet04} Stetson, P. B., McClure, R. D., \& VandenBerg, D. A.  2004, PASP, 116, 1012

\bibitem[sun (1991)]{sunt} Suntzeff, N. B.,\& Smith, V. V. 1991,  ApJ, 381, 160

\bibitem[Thoul et al. (1994)]{thoul} Thoul A., Bahcall J. \& Loeb A. 1994, ApJ, 421, 828

\bibitem[vandenberg (1985)]{vand}VandenBerg, D. A. 1985, in Proc. ESO Whorkshop 21, Production and Distribution of C, N, O Elements,
ed. I. J. Danziger, F. Matteucci, \& K. Kjar (Garching: ESO), 73 

\bibitem[vandenberg (2002)]{van2002}VandenBerg, D. A., Richard, O., Michaud, G., \& Richer, J. 2002, ApJ, 571, 487 

\bibitem[ven (2001)]{ventura} Ventura, P., D'Antona, F., Mazzitelli, I.,\& Gratton, R. 2001, ApJ, 550, L65

\bibitem[ven (2009)]{ventura09} Ventura, P., Caloi, V., D'Antona, F., Ferguson, J., Milone, A.,\& Piotto, G. P. 2009, MNRAS, 399, 934

\bibitem[zinn (1985)]{zin85} Zinn, R. 1985, ApJ, 293, 424


\end{thebibliography}
\end{document}